# The Need to Refine the Standard Recombination Theory and the Hubble Tension Problem

A. V. Shepelev[1,2,*]

[1]Department of Physics, I.M. Gubkin Russian State University, Moscow, Russia
[2]Scientific and Technological Centre of Unique Instrumentation of the Russian Academy of Sciences, Moscow, Russia

**Abstract** Over the past decades, cosmology has become largely based on experimental data, the most important sources of which are studies of the cosmic microwave background (CMB). CMB has been present in the Universe since the very first moments of its existence, and the features of CMB recorded now reflect the fundamental processes of the evolution of the Universe. These processes cause a weak anisotropy of CMB, which is no more than 0.01 percent. The structure of anisotropy, which is interpreted on the basis of a sufficiently detailed standard theory of primary recombination, allows us to establish the most important cosmological parameters. One of these parameters is the Hubble constant (more precisely, the Hubble parameter). However, the difference between the value of the Hubble constant, obtained from the results of the Planck mission, and obtained from independent local measurements, is 10%. Such a big difference is called the Hubble tension problem, and is an important problem of cosmology. The interpretation of the Planck measurements (as well as the interpretation of WMAP results, where a close value of the Hubble constant is obtained) is critically based on the predictions of the standard recombination theory. This work shows that the recombination rate is higher than predicted by standard theory and is caused by the excitation of the kinetic degrees of freedom of hydrogen atoms, which is ignored by standard theory. The calculations performed demonstrate that taking into account this process leads to a good coincidence of the values of the Hubble constant obtained from the results of the Planck experiment and local measurements.

## 1 Introduction

The Hubble tension, which is a high difference in the value of the Hubble constant according to the results of the Planck mission, 67.4+/-0.5 km/s/Mpc [1], and local measurements, 73.5+/-1.4 km/s/ Mpc [2], presents an intriguingly important problem of cosmology. Both of these values are obtained on the basis of objective observations, but the discrepancy between them exceeds $4.2\sigma$. To explain the Hubble tension, a number of variants of changing the standard ΛCDM model are proposed. However, if the results of objective observations contradict each other, it is necessary to check the correctness of the interpretation of the results of these observations, in particular, the interpretation of the results of the Planck experiment.

## 2. About the Speed of Primary Recombination

Interpretation of the results of the Planck mission critically depends[1] on the correctness of the theoretical

---

[1] When determining cosmological parameters based on the results of the Planck experiment [3], the authors explicitly note (quote): "Since the results of the Planck parameter analysis are crucially dependent on the accuracy of the recombination history, we have also checked, following Lewis et al. (2006), that there is no strong evidence for simple deviations

description of the primary recombination process – of the formation of hydrogen atoms from electrons and protons that make up the primary plasma. According to the standard recombination theory (SRT), its pace is slower than prescribed by the Saha law. Basically, SRT explains this by the slow dispersal of the 2p hydrogen state due to the high spectral energy density of the resonant Ly-alpha radiation that occurs during recombination. The fact is that Ly-alpha radiation, which is generated during the transition from the excited 2p state of the hydrogen atom to its ground 1s state, is immediately absorbed by other atoms in the 1s state, transferring them to the 2p state. This leads to an over-population of the 2p states of the atoms, which are then easily ionized by the equilibrium Planck background radiation, and as a result the atoms decay into protons and electrons. The density of Ly-alpha radiation is high, and their absorption is also high, so the rate of formation of atoms turns out to be low. According to SRT, the spectral energy density of Ly-alpha radiation is tens and hundreds of times higher than the spectral energy density of Planck's background at this frequency [4-6]. SRT explains such a high level of Ly-alpha radiation by the fact that it slowly comes out of interaction with hydrogen atoms, and this is almost exclusively due to the Hubble shift. Frequency diffusion of radiation was analyzed [7-10] using the Fokker-Planck operator. The authors postulated certain redistribution functions. However, heating of atoms by resonant radiation (excitation of external degrees of freedom of neutral hydrogen) and aberration of radiation during interaction with atoms are not taken into account in this approach. These processes can be taken into account by applying the Langevin equation for atoms; then there is no need to introduce a redistribution function. It is enough to consider thermodynamics and the exchange of pulses between radiation and atoms. This mechanism of Ly-alpha radiation escape, analyzed in detail earlier [11], is much faster than the escape due to the Hubble shift. In fact, this is the process of establishing a thermodynamic equilibrium between hotter radiation at the Ly-alpha frequency $2.467 \cdot 10^{15}$ Hz (the sum of Ly-alpha radiation and Planck radiation at this frequency) with the temperature $T_L$ [2] and colder atoms with the temperature $T_P$, which can be considered close to the temperature of the Planck's background [13].

Kinetics of the process of thermodynamic equilibrium in a system consisting of subsystems with different temperatures, in this case, with atoms and radiation at the Ly-alpha frequency, is established by an exchange of momentums of atoms and radiation[3] (SRT does not take into account this process). Similarly, to the Brownian process, the statistical nature of this interaction allows us to use the statistical differential Langevin equation for its description. The interaction between radiation and atoms has a discrete nature, therefore, the force acting on atoms is represented as the sum of the viscous force, since the atom is subjected to pressure from radiation, and the random force due to the discreteness of the interaction. As a result, we have the Langevin equation

$$\frac{dV}{dt} = -\xi V + f(t) \qquad (1)$$

where $f(t)$ is named the Langevin source. The solution of the Langevin equation is well known [14, 15] and gives the kinetics of establishing thermodynamic equilibrium in the atoms&radiation system [11]:

$$V^2(t) = V_0^2 e^{-2\xi t} + \frac{3kT_L}{m}\left(1 - e^{-2\xi t}\right) \qquad (2)$$

Here, $V(t)$ is an average velocity of the thermal motion of atoms, $V_0$ is an initial velocity. The parameter $\xi$ is proportional to the spectral energy density radiation at the Ly-alpha frequency $\rho_L$:

$$\xi = \frac{g_2}{g_1} \frac{c}{8\pi m v^2} A \rho_L \qquad (3)$$

The Einstein coefficient $A = \tau^{-1}$ is the probability of spontaneous transition of an atom from the excited state 2p to the ground state 1s, $\tau$ is the lifetime of the excited state, $m$ is mass of an atom, $g_2 / g_1$ is the ratio of the degeneracy of the levels 2p and 1s. As a result of Eq.(1), the bulk density of the kinetic thermal energy of the ensemble of hydrogen atoms increases with time as

$$W(t) = \frac{3}{2}nk(T_L - T_P) = \frac{3}{2}nk\Delta T\left(1 - e^{-\frac{t}{t_r}}\right)$$

$$t_r = \left(\frac{g_2}{g_1}\frac{c}{4\pi m v^2}A\rho_L\right)^{-1} \qquad (4)$$

(Here $n$ is the concentration of atoms). The energy balance condition in the system of atoms and radiation at the Ly-alpha frequency gives an upper estimate of the time at which the difference between the temperatures $T_L$ and $T_P$ decreases by half [11]:

$$\Delta t = \frac{128\pi}{3}\ln 2 \cdot \frac{g_1}{g_2} \cdot \frac{v^3}{c^3}\frac{T_p}{\Delta T}(nA)^{-1} = \frac{128\pi}{3}\ln 2 \frac{g_1}{g_2}\frac{T_p}{\Delta T}\frac{\tau}{\lambda^3 n} \qquad (5)$$

---

from the assumed history. However, we note that any deviation from the assumed history could significantly shift parameters compared to the results presented here and we have not performed a detailed sensitivity analysis."

[2] The temperature of radiation at a certain frequency is uniquely related to the spectral brightness of the radiation. In the case of isotropic radiation, the spectral brightness is proportional to the spectral energy density [12].

[3] For a very simple rough estimate using the 'photonic language', see Appendix. This approach is visual, but very inaccurate. 'Photon language' is not used in calculations in the main text.

**Table 1.** $\eta$ and $\Delta z_L$ at different ratios $\rho_L / \rho_P$

| $\rho_L / \rho_P$ | 1.1 | 1.5 | 2.0 | 5.0 | 10 | 50 |
|---|---|---|---|---|---|---|
| $\eta$ | 2.0 | 2.2 | 2.4 | 3.1 | 3.9 | 7.0 |
| $\Delta z_L$ | $3.40 \cdot 10^{-3}$ | $8.0 \cdot 10^{-4}$ | $4.6 \cdot 10^{-4}$ | $2.0 \cdot 10^{-4}$ | $1.3 \cdot 10^{-4}$ | $0.7 \cdot 10^{-4}$ |

In order to establish a decrease not in temperature, but in the spectral energy density at the Ly-alpha frequency, it is necessary to use a universal relationship between the spectral energy density and the radiation temperature

$$\rho_L = \frac{8\pi h \nu^3}{c^3} \frac{1}{\exp\frac{h\nu}{kT_L} - 1}, \quad \rho_P = \frac{8\pi h \nu^3}{c^3} \frac{1}{\exp\frac{h\nu}{kT_P} - 1}. \quad (6)$$

This allows us to estimate the value of

$$\eta = \frac{\rho_L - \rho_P}{\rho_{L\Delta t} - \rho_P} \quad (7)$$

This value shows how many times the difference between the spectral energy densities of radiation at the Ly-alpha frequency and of the Planck background is weakened during the time $\Delta t$ determined by Eq.(4); the concentration of hydrogen atoms $n$ is assumed to be $0.3 \times 10^9$ m$^{-3}$, the same as in [11]. Here, for convenience, $\Delta t$ is expressed in terms of the redshift change $\Delta z_L$. The values of $\Delta z_L$ for $z = 1300$ are given in Table 1 (for more information, see [16]).

Redshift change $\Delta z_L \approx 4.6 \cdot 10^{-4}$ if the spectral density of Ly-alpha radiation is equal to the spectral density of the Planck background at this frequency, and $\Delta z_L \approx 2.0 \cdot 10^{-4}$ if it exceeds the Planck background four times (calculation details in [16]). Such redshift change is 2-3 orders of magnitude less than the redshift change due to Hubble shift $\Delta z_H = \Delta \nu \cdot z / \nu \approx 5.5 \cdot 10^{-2}$.

Thus, resonant Ly-alpha photons quickly reduce their energy by decreasing the frequency, and get out of interaction with atoms. Therefore, they do not accumulate, there are fewer of them in the orders of magnitude than SRT predicts, and they do not significantly affect the rate of recombination through the 2p state.

## 3. The Sound Horizon of the Recombination Epoch and The Hubble Constant

The higher the recombination rate, the greater the redshift $z_*$ corresponding to the last scattering; according to SRT, the value of $z_*$ is approximately 1090. In a number of works (see, for example, [17] and [18]), the answer is given to the question of what happens when the Lyman alpha radiation is "disabled" while preserving all other currently known processes: if Lyman-alpha radiation is neglected, the recombination is described by Saha's law. This changes a redshift corresponding to the last scattering.

This value is usually obtained by constructing the visibility function $g(z)$,

$$g(z) = \frac{d\tau(z)}{dz} \cdot \exp(-\tau(z)) \quad (8)$$

and finding its maximum. Here,

$$\tau(z) = \int \frac{\sigma_T c n_e}{(1+z)H(z)} dz \quad (9)$$

is the optical depth, $\sigma_T$ is the Thomson scattering cross section, $n_e$ is an electron density [6]. But the redshift corresponding to the last scattering can be found without searching a maximum of the visibility function $g(z)$, since the condition of the maximum $\frac{dg(z)}{dz} = 0$ leads to an algebraic equation

$$\frac{d\tau(z)}{dz} - \tau^2(z) = 0 \quad (10)$$

The electron dencity $n_e(z)$ is

$$n_e(z) = n_0 x_e (1+z)^3 \quad (11)$$

If we assume that the relative concentration $x_e$ is described in accordance with the Saha equation ([19], [20]), the numerical solution of Eq.(10) allows us to obtain the redshift $z_{*1}$ corresponding to the last scattering according to this recombination model:

$$z_{*1} = 1279 \quad (12)$$

The magnitude of the redshift of the last scattering makes it possible to determine the sound horizon $r$ of the recombination epoch, which is further considered as an independent parameter, see [21]:

$$r = \int_z^\infty \frac{c_s}{H(z)} dz \quad (13)$$

Here,

$$c_s = \frac{c}{\sqrt{3(1+R_B)}} \qquad (14)$$

is a speed of sound, $R_B = 0.76\rho_b / \rho_r = 0.45$,

$$H(z) = H_0\sqrt{\Omega_r(1+z)^4 + \Omega_m(1+z)^3 + 1 - \Omega_r - \Omega_m} \qquad (15)$$

Determining the ratio of the sound horizon of the recombination epoch to the commoving distance from a present day observed to the last scattering distance $D(z)$ is one of the most important results of the Planck experiment and determines the angular scale $\theta$ of the sound horizon of the recombination epoch:

$$\theta = \frac{r}{D(z)} = 0.01041. \qquad (16)$$

It is this value that is experimentally determined from the angular power spectrum of CMB temperature fluctuations obtained in the Planck experiment.

It is easy to estimate the magnitude of the sound horizon $r_{*1}$ corresponding to the redshift $z_{*1}$, knowing the magnitude of the sound horizon $r_*$ corresponding to the $z_*$ predicted STR. For large redshift values

$$\int_z^\infty \frac{c_s}{H(z)} dz \approx \int_z^\infty \frac{c_s}{H_0\sqrt{\Omega_r(1+z)^4 + \Omega_m(1+z)^3}} dz \qquad (17)$$

which yields the estimate of the $r_{*1}$ value:

$$r_{*1} = \frac{r_{*1}}{r_*} r_* \approx \frac{\sqrt{\frac{\Omega_m}{z_{*1}} + \Omega_r}}{\sqrt{\frac{\Omega_m}{z_*} + \Omega_r}} r_* = \sqrt{\frac{z_*}{z_{*1}}} r_* = 0.923 r_* \qquad (18)$$

The value of the sound horizon of the recombination epoch allows us to determine the value of the comoving sound horizon at the end of the drag epoch $r_d$, which turns out to be approximately two percent higher than the value of the sound horizon. The value $r_*$ corresponds to $r_{d*} = 1.02 r_* = 147$ Mpc [1].

Equation (18) gives an estimate of the comoving sound horizon at the end of the drang epoch $r_{d*1}$, obtained taking into account the excitation of kinetic degrees of freedom of hydrogen atoms during recombination:

$$r_{d*1} \approx 0.923 \cdot r_{d*} \approx 136 \text{ Mpc} \qquad (19)$$

Decreasing the sound horizon at recombination inevitably leads to an increase in the value of the Hubble constant [21]. As shown by the authors, the value $r_{d*1} = 136$ Mpc gives a value of $H_0$ the order of 72-74.5 km/s/Mpc. This is in agreement with the definition of the Hubble constant based on local measurements, 73.5 km/s/ Mpc [1].

## 4. Conclusions

The value of $H_0$ is obtained here based on the solution of the Langevin equation and the conditions of thermodynamic balance. To achieve greater accuracy, the process of interaction of hydrogen atoms and radiation must be calculated more adequately than before. The population dynamics calculation used in SRT, based on the balance equations, is actually just an approximation of the kinetic equation method for the components of the density matrix developed in quantum electronics, and does not take into account the radiation relaxation process. In kinetic equations for the components of the density matrix, relaxation processes in a complex system (atoms, equilibrium Planck radiation, nonequilibrium recombination radiation) should be taken into account, although this will not be a simple matter, since radiation also relaxes and cannot be considered as a thermostat in principle.

However, this is necessary to clarify the theory of recombination, in particular for an adequate interpretation of the results of the Planck mission.

## Acknowledgements

The author is grateful to G.Kh. Kitaeva, M.V. Milovidov and N.N. Rukavishnikov for their help in preparing this work.

## Data Availability

The data that support the findings of this study are available within this paper and from the author upon reasonable request.

## Appendix

Having absorbed a photon, the atom transits into an excited state and then returns to the initial state emitting a phonon. If we assume that the atom is initially motionless, according to the law of momentum conservation

$$\mathbf{p}_{in} = \mathbf{p}_{em} + \mathbf{p}_{atom} \qquad (A1)$$

Here, $\mathbf{p}_{in}$, $\mathbf{p}_{em}$, and $\mathbf{p}_{atom}$ are the momentums of the incident photon, the emitted phonon, and the atom, correspondingly. Therefore,

$$p_{atom}^2 = p_{in}^2 + p_{em}^2 - 2\mathbf{p}_{in}\mathbf{p}_{em} \qquad (A2)$$

The product $\mathbf{p}_{in}\mathbf{p}_{em}$ is zero on average, since the directions of the vectors $\mathbf{p}_{in}$ and $\mathbf{p}_{em}$ are independent. Since the atom is massive, $p_{em} \approx p_{in}$. As a result, an atom with mass $m$ gains an average momentum $p_{atom} = \sqrt{2}p_{in} = \sqrt{2}\frac{h}{\lambda}$ and energy

$$\frac{1}{2m}p_{atom}^2 = \frac{1}{2m}\left(\frac{\sqrt{2}h}{\lambda}\right)^2 = \frac{h^2}{m\lambda^2} \qquad (A3)$$

This energy arises due to a small decrease in the photon frequency. After $N$ collisions of a photon with atoms, the photon will leave half the energy required to exit the Doppler band $\Delta\nu$. So,

$$N = \frac{1}{2}h\Delta\nu \Big/ \frac{h^2}{m\lambda^2} = \frac{mc\Delta\lambda}{2h}. \qquad (A4)$$

For another hand, the number of collisions of a photon with atoms during a time $\Delta\tau$ is

$$N = \sigma c n \Delta\tau \qquad (A5)$$

where $\sigma$ is the absorption cross section of Ly-alpha radiation, $n$ is concentration of hydrogen atoms. So, the photon will leave the Doppler band, i.e. will practically stop interacting with hydrogen atoms, within a time

$$\Delta\tau = \frac{m\Delta\lambda}{2\sigma n h} \qquad (A6)$$

Assuming for a numerical estimate that the temperature of atoms is $T_p = 3550 K$, and $n = 0.3 \cdot 10^9 \, \text{m}^{-3}$, we obtain $\Delta\tau \approx 3.4 \cdot 10^6$ s, which corresponds to a redshift change $\Delta z \approx 4.5 \cdot 10^{-4}$. This is a rough estimate, since the motion of atoms and thermodynamics are not taken into account. Correct calculation taking into account temperature gives a close value, see the main text.